\documentclass[default,iicol]{sn-jnl}

\usepackage{graphicx}
\usepackage{changepage}
\usepackage{lipsum}
\usepackage{xcolor}
\usepackage{amssymb}
\usepackage{amsmath}
\usepackage{mathrsfs}
\usepackage[normalem]{ulem}  
\usepackage{url}
\usepackage{tabularx}
\usepackage{adjustbox}
\usepackage{booktabs}
\usepackage{array}
\usepackage{caption} 

\jyear{2021}%

\theoremstyle{thmstyleone}%
\theoremstyle{thmstyletwo}%

\theoremstyle{thmstylethree}%

\usepackage[ampersand]{easylist}
\newcolumntype{?}{!{\vrule width 1pt}}

\makeatletter
\def\thickhline{%
  \noalign{\ifnum0=`}\fi\hrule \@height \thickarrayrulewidth \futurelet
  \reserved@a\@xthickhline}
\def\@xthickhline{\ifx\reserved@a\thickhline
              \vskip\doublerulesep
              \vskip-\thickarrayrulewidth
             \fi
      \ifnum0=`{\fi}}
\makeatother

\newlength{\thickarrayrulewidth}
\begin{document}

\title[Generalizing DLA]{Generalizability in Document Layout Analysis for Scientific Article Figure \& Caption Extraction}

\author*[1,2]{\fnm{Jill} \sur{Naiman}}\email{jnaiman@illinois.edu}

\affil*[1]{\orgdiv{School of Information Sciences}, \orgname{University of Illinois, Urbana-Champaign}, \orgaddress{\street{Daniel ST}, \city{Champaign}, \postcode{61820}, \state{Illinois}, \country{USA}}}

\affil[2]{\orgdiv{National Center for Supercomputing Applications}, \orgname{University of Illinois, Urbana-Champaign}, \orgaddress{\street{W Clark ST}, \city{Urbana}, \postcode{61801}, \state{Illinois}, \country{USA}}}


\abstract{ 
The lack of generalizability -- in which a model trained on one dataset cannot provide accurate results for a different dataset -- is a known problem in the field of document layout analysis.
Thus, when a model is used to locate important page objects in scientific literature such as figures, tables, captions, and math formulas, the model often cannot be applied successfully to new domains.
While several solutions have been proposed, including newer and updated deep learning models, larger hand-annotated datasets, and the generation of large synthetic datasets, so far there is no ``magic bullet" for translating a model trained on a particular domain or historical time period to a new field.
Here we present our ongoing work in translating our document layout analysis model from the historical astrophysical literature to the larger corpus of scientific documents within the HathiTrust U.S. Federal Documents collection.
We use this example as an avenue to highlight some of the problems with generalizability in the document layout analysis community and discuss several challenges and possible solutions to address these issues.
All code for this work is available on The Reading Time Machine GitHub repository, \url{https://github.com/ReadingTimeMachine/htrc_short_conf}.
}



\maketitle

\section{Introduction}

Much of the science communicated through academic literature is transmitted through page components other than pure text -- namely, figures, tables and mathematical formulas.
Thus, these kinds of document objects are of particular interest to scientists and those who study science communities and processes 
\citep{maltese_data_2015}.

While newer academic articles are ``born-digital", meaning the literature is stored in formats that make page objects easy to extract (e.g. as in XML), this is not true for articles that were published prior to the wide distribution and access capabilities of the internet \citep{adsass2012}.
 This ease of page object extraction can sometimes extend to vector-based, i.e. ``rule-based" PDFs whose file types contain the instructions for rendering article pages.  
If the vector-PDF format is known, then text and images can potentially be extracted with heuristics which search for keywords \cite{Choudhury2013,pdffigures2} and tables/figures and their captions can be extracted as pairs \citep{pdffigures2,GROBID}.  However, heuristic extraction and indexing of figures from  vector-based PDF documents can often be non-trivial leading to erroneous or missing page objects \citep{surveydeeplearning}.

More recently, deep learning methods have typically been employed to extract page objects from born-digital and scanned documents using a variety of methods \citep{surveydeeplearning,saha2019,yang2017layout}.
In the cases of born-digital documents, these deep learning methods are combined with heuristically-derived results in post-processing steps \citep{deepfigures}.
However, for historical scanned documents, these methods present many challenges \citep{scanbank,scanbankthesis,naiman2022}.

Given that HathiTrust has a wealth of these kinds of data products available to the researcher, it is worth discussing some of the issues that are inherent in attempting to extract and categorize these page objects.
In what follows, we focus on one major issue in the document layout analysis community, namely the difficulties with {\textit{generalizability}} of both heuristic and deep learning models -- when a model is trained on a particular type of page (i.e. electronic thesis and dissertations) it does not tend to perform well on another type of page (i.e. academic articles).

\section{The Problem of Generalizability}

The lack of generalizability is a known and pervasive problem in the field of document layout analysis \citep[e.g.][]{surveydeeplearning,doclaynet}.  
A change in not only publication type, but simply publication year can drastically lower the accuracy of page object extraction methods for models that are not explicitly trained on this type of document \citep{scanbank,naiman2022,naimanprep}.

Our prior work was aimed at the extraction of figures and their captions from a subset of the ``pre-digital" astrophysical literature holdings of the Astrophysics Data System (ADS)\footnote{\url{https://ui.adsabs.harvard.edu/}} using both grayscale and optical character recognition (OCR) features of article pages \citep{naiman2022}.
Our model produced a high level of accuracy on our dataset  -- for an intersection-over-union (IOU) metric of 0.9 we found F1 scores of $\geq 90\%$\footnote{The F1 score is a combination metric of precision (prec) and recall (rec) with $\rm{F1 = 2\times prec \times rec/(prec+rec)}$ with $\rm{prec = TP/(TP+FP)}$ and $\rm{rec = TP/(TP+FN)}$ as the combination metrics of true positives (TP), false positives (FP) and false negatives (FN). }\citep{naiman2022,naimanprep}.

\begin{figure}[!htp]
\centering
\includegraphics[width=0.5\textwidth]{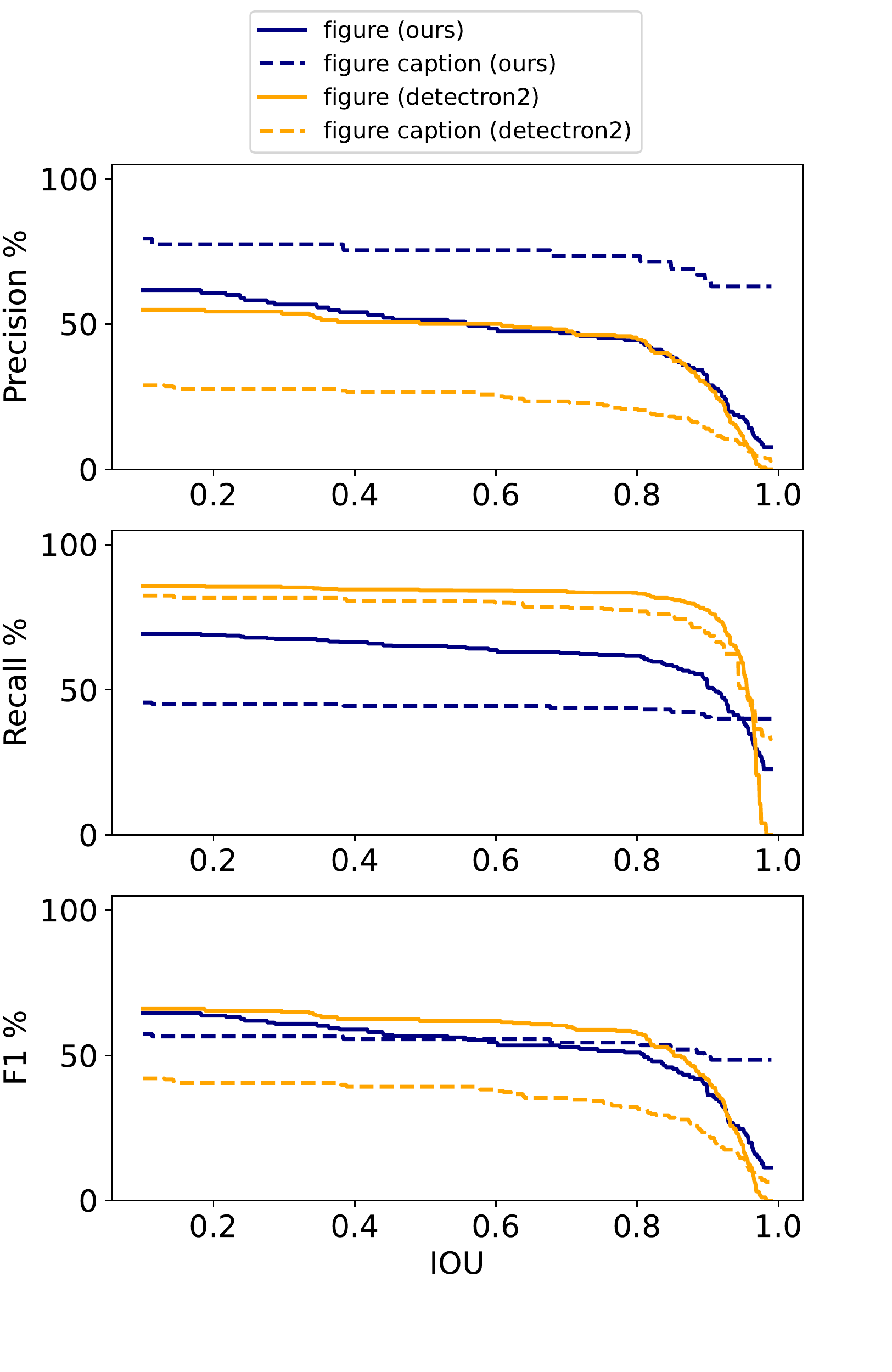}
\caption{Precision (\textit{top}), Recall (\textit{middle}) and F1 score (\textit{bottom}) at different IOU cut-offs for our model  \citep{naiman2022,naimanprep} (blue lines) and \textsf{detectron2} \citep{wu2019detectron2} trained on the PubLayNet document dataset \citep{publaynet} (orange lines). Neither model generalizes well to the HathiTrust set of documents. As shown in the F1 score in the bottom panel, \textsf{detectron2} performs better with figures (solid lines) at low IOU cut-offs, and our model better with captions (dashed lines) and both figures and captions at higher IOU cut-offs.  Metrics are calculated from 363 randomly chosen pages from six HathiTrust articles (see text for more details).}
\label{fig:metrics}
\end{figure}

A natural extension of our work would be the extraction of figures and their captions in other scientific fields or journals.  As the HathiTrust contains potentially millions of such article pages a possible fruitful endeavor would be the application of our model to such articles.
To illustrate some of the issues an application of our model to the HathiTrust corpus we apply our model to randomly selected pages from articles within the U.S. Federal Documents\footnote{\url{https://babel.hathitrust.org/cgi/mb?a=listis&c=2062901859}} collection.  To give our model the best chance of success in our comparison, we subset this collection with the search fields of ``astronomy" and filter the results to include only English-language Conference, Journal and Manuscript documents with ``Full View" available, bringing the total records to 56,282.
For our illustration, we select $\approx$350 randomly chosen pages from six articles within this filtered collection which we annotate with figure and caption class definitions from \citep{naiman2022,naimanprep} in MakeSense.ai\footnote{\url{https://www.makesense.ai/}}\citep{makesense}.
The precision, recall and F1 score at different IOU cut-offs for our model applied to this data are shown by solid and dashed navy blue lines in \autoref{fig:metrics}.
As one can see -- these lines clearly lie below the F1$\approx$90\% we found with our original dataset.

One possible explanation is simply that our dataset of $\approx$6000 annotated pages, while large enough of a dataset to train a YOLO-based model like ours \cite{bochkovskiy2020yolov4,Wang_2021_CVPR} is too small to create a model that generalizes well.
To test this possibility, we also apply a version\footnote{\url{https://github.com/JPLeoRX/detectron2-publaynet}} of Facebook's \textsf{detectron2} \citep{wu2019detectron2} that has been trained on the PubLayNet\footnote{\url{https://github.com/ibm-aur-nlp/PubLayNet}} dataset \cite{publaynet}.
As the PubLayNet dataset contains over 1 million articles, this should provide a sufficiently large number of training cases for \textsf{detectron2} to generalize well, if it is indeed the size of the training set that is affecting the model's generalizability. 
However, as shown by the solid and dashed orange lines of \autoref{fig:metrics}, this is not the case.
While the F1 score for the \textsf{detectron2} model is higher than ours for figures, it does less well for figure captions (though, as discussed in \citep{naiman2022} this may be due to inconsistencies between ``caption" and ``text" classes, the latter being the only class trained in this version of \textsf{detectron2}).
Additionally, the metrics decrease slightly more sharply at high IOU levels for the \textsf{detectron2} model when compared to ours.  
As motivated in \citep{naimanprep} -- when using object detection methods like our YOLO-based model and the Faster/Mask-RCNN models used in \textsf{detectron2}, maintaining high accuracy at large IOU cut-offs is vital for being able to ``extract" page objects from the page and retain their readability and overall usefulness.
Thus the drop off in accuracy at high IOU for both models is a concern.

\begin{figure*}[!htp]
\centering
\includegraphics[width=0.99\textwidth]{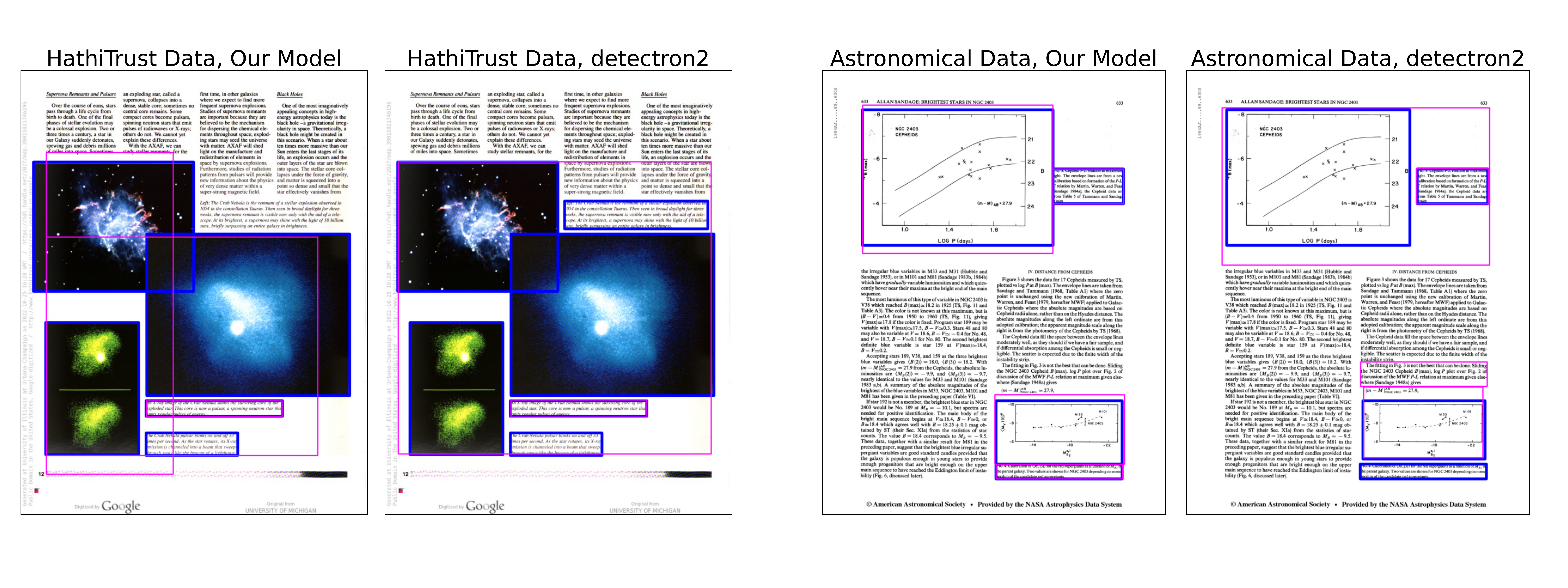}
\caption{Ground truth boxes (blue) and boxes found (magenta) from both the model presented in \citep{naiman2022} and the \textsf{detectron2} model \citep{wu2019detectron2} trained on the PubLayNet dataset \citep{publaynet}.  While both models do not do well on an example HathiTrust page (\textit{left two panels}), they are more accurate when applied to page objects from a page in the astrophysical dataset from ADS (\textit{right two panels}).}
\label{fig:qual}
\end{figure*}

One possibility for the significant decrease in accuracy for both models is simply that the data they have been trained with is too ``different" from the new HathiTrust data to which they are now being applied. 
This suggestion is depicted qualitatively in \autoref{fig:qual} which shows how both our model and \textsf{detectron2} fair on a page from the HathiTrust dataset (left two panels) and our original astrophysical dataset from ADS (right two panels).
Both our model and \textsf{detectron2} miss the unfamiliar layout elements such as the overlapping figures and ``floating" figure captions present in the HathiTrust page.

Our model does best on our dataset and understandably so -- it has been trained specifically on astrophysical literature while the \textsf{detectron2} model has been trained on medical scientific literature from PubMed \citep{publaynet}.

\autoref{fig:qual} highlights the differences in accuracies that can be present for different types of documents -- in this case the difference in accuracies between ``Manhattan-style" (academic) documents and ``non-Manhattan-style" (non-academic) documents.
Further, the failure of \textsf{detectron2} to accurately find boxes in the astrophysical dataset (right-most plot of \autoref{fig:qual}) highlights how sensitive different models are to changes in the layout of document pages.

\section{The Answer is Always More Data}

Given that seemingly slight differences in font size and page object spacing can result in drastic drops in accuracy of machine learning models, one natural solution might be to increase the size and scope of the datasets used in training.
While simple in theory, in practice one can run across several issues of concern.

\subsection{Inconsistent Page Object Definitions}

Before attempting a large scale data collection initiative one must define what it is precisely they are looking for on a document's page.
As discussed in \citep{naimanprep}, different datasets are often annotated with different definitions of page objects such as figures and figure captions.  For example, sometimes figure captions are included in figure box definitions, and sometimes each panel of a figure is annotated as a figure, while in other cases all panels of a figure are considered inside a figure box (see Figure 2 of \citep{naimanprep}).

When different annotators work on the same dataset, they can disagree on the class definitions leading to inconsistent data \textit{within the same dataset} \citep{younas2019,doclaynet}.
While there have been pushes to adopt a standardized methodology for defining page object classes \citep[e.g.][]{pletschacher_page_2010}, these have yet to be fully adopted by the different communities involved with document layout analysis.

Additionally, the definitions of these different classes may \textit{depend on use case}.  In the work presented in \citep{naiman2022,naimanprep}, we create the class definitions of figure and figure caption based on our ultimate goal -- hosting historical figures and their captions on the Astronomy Explorer (AIE)\footnote{\url{http://www.astroexplorer.org/}} as is automatically done with newly published articles.
Thus, even if consistent annotation ``codebooks" can be developed (as was done in \citep{naiman2022,naimanprep}) these definitions may not suit the needs of all document layout analysis applications.

\subsection{Expensive Human Annotation}

Assuming there is agreement within a group or field as to the definitions of page object classes, one option for generating data is to employ a large group of annotators to hand-classify article pages.
One such effort is the DocLayNet dataset \citep{doclaynet}.
This large scale annotation effort illustrates many of the logistic and resource requirements for the generation of a large and diverse set of documents for use with machine learning models.

The DocLayNet dataset, which contains over 80,000 manually annotated pages, took $\sim$40 supervised annotators and a small group of experts over 6 months to produce, with only a ``small fraction" of the pages being seen by more than one annotator \citep{doclaynet}.
Additionally, while the annotation process included a $>$100 page annotation guide and a 12 week training period, the resulting intercoder agreement was $\approx$80-85\% for their 11 page object categories in the ``small fraction" of double and triple-annotated pages \citep{doclaynet}. 

The DocLayNet dataset represents arguably one of the state-of-the art human-annotated datasets for document layout analysis.
The significant logistics and time requirements for the generation of this dataset, and its resulting intercoder agreement of 80-85\% for each class demonstrates how difficult these hand-annotation tasks are to scale.
Relevant to our work, this dataset does not include scanned pages -- the DocLayNet authors point out these pages are often warped or non-uniformly colored introducing even more possible uncertainty and errors into the annotation process \citep{doclaynet}.

\subsection{Minimal Historical Synthetic Data}

Given the time requirements for producing datasets like DocLayNet, another option is to generate ``synthetic data" which can be built from article source files.

Large ``benchmark" datasets have been created by mining XML \citep[e.g. PubLayNet][]{publaynet} and LaTeX source files with or without weak supervision \citep[e.g. TableBank and DocBank][]{tablebank,docbank}.
However, the majority of articles included in such datasets are published recently -- for example only $\approx$3\% of the PubLayNet dataset is comprised of articles older than 1997.

Given this dearth of historical synthetic datasets, there have been several efforts to artificially ``age" these newer documents by including effects such as artificial warping, rotation, and simulated dust and random noise on the page.
While much focus has been placed on the aging of articles on downstream tasks such as the mining of historical event-related OCR text \citep{eventmining} or named-entity recognition \citep{namedentityrecog}, some recent work has focused on the effects of the aging process on the localization of page-objects \citep{scanbank,scanbankthesis} and the generation of new training sets for historical documents \citep{synthmod}.   

Synthetic historical data is certainly a promising avenue for increasing training dataset sizes, however many datasets are still relatively small and constructed for a particular type of historical document \citep[e.g. predominantly books as in][]{monnier2020docExtractor}.

\section{...Unless the answer is better models?}

As there are issues in training data collection and generation, another path to investigate is the creation of better models to use with the limited training data available.

One possibility is the use of models that have been trained on large document layout analysis or object detection datasets that are ``fine-tuned" to a particular historical document set through transfer learning.  
There are several document layout analysis works that adopt this strategy \citep{boukhers_mexpub_2021,schreiber_deepdesrt_2017,tabletransferlearning}.
However, for historical documents, there is some evidence that transfer learning may add little to page object localization accuracy \citep{scanbankthesis}.
Additionally, this requires the format of each page to be the same, making the addition of novel features to the training process more difficult to incorporate \citep{naiman2022}.

Beyond transfer learning, models that move beyond translating object detection to document layout analysis tasks are those that include the text data as training features \citep{younas2019,naiman2022,naimanprep}.  Often these models are ``multi-modal" in that they draw from the fields of machine learning methods for image classification and segmentation and the processing of text with natural language processing or similar techniques \citep{boukhers_vision_2022}.

Additionally, models are being developed that explicitly deal with a dearth of training data.  These ``one-shot" or ``few-shot" models have the potential to drastically decrease the amount of time and logistics that must be resourced creating a specialized document dataset for a particular field \citep{singh_multi-domain_2020}.

The combination of models that can make use of transfer learning, textual document components and only require a small number of training instances could potentially drive down the ``cost" of analysing the historical literature of a particular field.  However, at present, most of the state-of-the art models have difficulty meeting the high levels of accuracy that are present in other object detection applications \citep{doclaynet,surveydeeplearning}.

\begin{figure*}[!htp]
\centering
\includegraphics[width=0.99\textwidth]{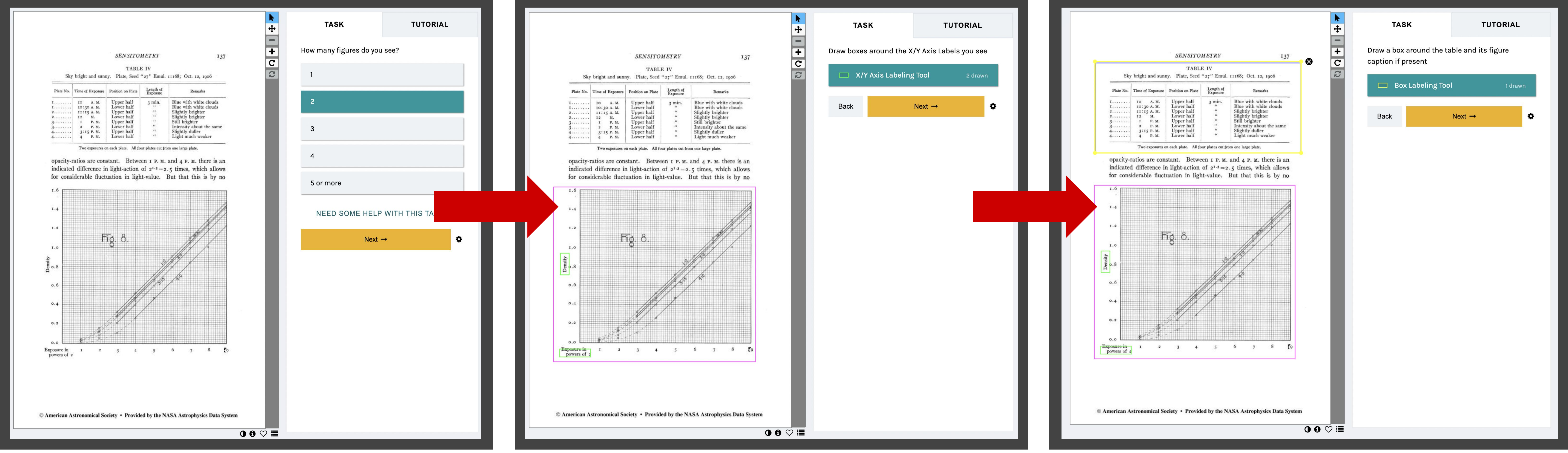}
\caption{Example of several steps of a prototype page object detection interface on the citizen science platform Zooniverse which takes an annotator from identifying the number of figures on a page (\textit{left}) through drawing boxes around figures and their axis labels (\textit{middle}) and identifying other interesting page objects like tables (\textit{right}). The goal of our interface design is to efficiently guide a citizen science partner through an annotation of a scientific article that will result in high levels of intercoder agreement and be implementable at scale.}
\label{fig:zoonie}
\end{figure*}

\section{Just kidding!  Of course the \textit{real} answer is All The Things!}

Given that there is presently no ``magic bullet" to perfectly localize page objects across diverse domains, it is likely the solution needed to increase the accuracy of document layout analysis will be a combination of the multiple approaches discussed here into a \textit{process}.

In our own work on translating our model tuned on the astrophysical literature to the wider historical scientific corpus within the HathiTrust, we are approaching the generalization problem with three major prongs: \textit{document-specific machine learning models, citizen science to scale the annotation process, and the generation of large synthetic datasets with appropriate page-aging processes applied.}

Our current model detects figures and their captions with a modified object detection model which makes use of both OCR, linguistic, and grayscale features \citep{naiman2022,naimanprep}.  Given its architecture (YOLO-based), it has a low false positive rate, however we plan to combine it in an ensemble with models like \textsf{detectron2} (Mask/Faster-RCNN-based) which have a different error profile, thereby likely increasing our model accuracy.  Additionally, we plan to investigate transfer learning options such as few-shot models \citep[e.g.][]{singh_multi-domain_2020} to decrease the number of training instances required to fine-tune our model to new domains.

To create a curated gold-standard dataset, we are currently designing an annotation interface on the Zooniverse\footnote{\url{https://www.zooniverse.org/}} platform -- a prototype of which is shown in \autoref{fig:zoonie}.
The Zooniverse platform hosts over one million citizen scientists who help professional scientists with their work.  
The platform allows for many options to help facilitate the scaffolding required for a successful partnership with citizen scientists in creating accurate annotations.  Through tools like tutorials, talk forums and the ability to help citizen scientists ``test" their knowledge with gold-standard datasets \citep{tess}, much of the work to train annotators can be performed at scale.
With a larger group of people performing annotations, metrics like intercoder agreement and tolerance for machine learning models can be more fully quantified \citep{zoonie1,zoonie2,zoonie3}.

Finally, we are currently studying the aging process of scientific article documents in more detail.  While this work is preliminary and ongoing, we plan to make use of synthetic datasets in order to train our models once the quantification of the page aging process is complete.

While past work has typically focused on one of the main methods of increasing document layout analysis accuracy (models, curated datasets, and synthetic data), recent work from the community has illustrated the importance of combining all three.  Thus our ongoing work is aimed at publishing not only the results of our efforts, but the processes we have used to guide our thinking with the hope that it proves useful for others in the document layout analysis community.

This work is supported by a Fiddler Fellowship and a NASA Astrophysics Data Analysis Program Grant (20-ADAP20-0225).

\bibliographystyle{sn-mathphys}
\bibliography{references}



\end{document}